# Regulating Reality: Exploring Synthetic Media Through Multistakeholder AI Governance

Claire R. Leibowicz

Oxford Internet Institute, University of Oxford, claire.leibowicz@oii.ox.ac.uk

Artificial intelligence's integration into daily life has brought with it a reckoning on the role such technology plays in society and the varied stakeholders who should shape its governance. This is particularly relevant for the governance of AI-generated media, or synthetic media, an emergent visual technology that impacts how people interpret online content and perceive media as records of reality. Studying the stakeholders affecting synthetic media governance is vital to assessing safeguards that help audiences make sense of content in the AI age; yet there is little qualitative research about how key actors from civil society, industry, media, and policy collaborate to conceptualize, develop, and implement such practices. This paper addresses this gap by analyzing 23 in-depth, semi-structured interviews with stakeholders governing synthetic media from across sectors alongside two real-world cases of multistakeholder synthetic media governance. Inductive coding reveals key themes affecting synthetic media governance, including how temporal perspectives—spanning past, present, and future—mediate stakeholder decision-making and rulemaking on synthetic media. Analysis also reveals the critical role of trust, both among stakeholders and between audiences and interventions, as well as the limitations of technical transparency measures like AI labels for supporting effective synthetic media governance. These findings not only inform the evidence-based design of synthetic media policy that serves audiences encountering content, but they also contribute to the literature on multistakeholder AI governance overall through rare insight into real world examples of such processes.

## 1 INTRODUCTION

For centuries, visual technologies have enabled new ways of depicting and understanding the world. As Kelsey and Roberts [26] suggest, visual technologies "extend beyond [their] typical wires-and-switches connotation to [include] any innovation that has extended, structured, or transformed visual perception and communication." Synthetic media, also known as generative or AI-generated media, is one such example of a modern visual technology that implicates vast societal dynamics, ranging from child safety to politics to artistic expression [52, 3]. I adapt the 2023 U.S. Executive Order on AI's definition of synthetic content to refer to synthetic media as "images, videos, and audio clips, that [have] been significantly altered or generated by algorithms, including by AI" [8]. Partnership on AI's definition of synthetic media helps specify what "significant alteration or generation" means, suggesting such content is "highly realistic, would not be identifiable as synthetic to the average person, and may simulate artifacts, persons, or events" [40].

Synthetic media's realism has drastically improved over the past five years, enabled by advancements in machine learning [53]. While such advancements support new opportunities for responsible uses like storytelling and privacy preservation [51, 45], they also accelerate deception, defamation, and harassment [31]. Beyond realism, several dynamics motivate scholarly attention on synthetic media, and by extension, how the technology should be governed. In the past three years, anchored on the launch of OpenAI's Dall-E, text-to-image generators have made it increasingly easy and fast for those lacking technical skills to generate visuals with AI [19]. Such tools, aided by computational and image recognition improvements, meaningfully shifted the socio-economic calculus and impact of creating visuals with AI [43, 54, 16]. Meanwhile, the primacy of visuals in daily life intensified; today, people consume their news through YouTube videos, learn about history through TikTok reels, and even decide whom to date based on digital images. Based on a 2024 report from OfCom, UK teenagers relied more readily on YouTube, TikTok, and Instagram to get their news—all of which are largely visual social media platforms—than text-based social media like WhatsApp and X, and traditional news outlets like the BBC [38]. Indeed, digital visual experience has taken shape as the dominant way for individuals to understand the world, at the very moment that AI advancements have made it easier to synthesize and manipulate such visual experiences.

Many discrete stakeholder groups are interested in the implications of these visual trends. For example, technology platforms are concerned about the capacity of synthetic videos to manipulate users before elections or buy fake items, yet are excited to see how users can express themselves in new ways with the technology; journalists are worried about imposters creating synthetic videos that purport to come from their trusted news brand while also using the technology to preserve anonymity of their sources; human rights defenders focus on the ways synthetic media may render videos as evidence of abuses useless while enabling creative ways to speak truth to power; and artists are excited about the opportunities to express themselves through synthetic media, while also being concerned about how their creations might be leveraged without their consent to train AI [55, 18, 20]. It is not only the largest technology companies that often feature the greatest resources for synthetic media development that affect synthetic media; media, civil society, and other stakeholders are also integral to understanding and shaping synthetic media's impacts [29].

### 1.1 Synthetic Media Governance: Norms, Standards, Frameworks, and More

Synthetic media's wide-ranging societal implications have led to increased attention on the governance of such technology. Here, I describe AI governance broadly to include policies, norms, standards, regulations, and technological interventions both within and beyond government [10]. Efforts to govern synthetic media crescendoed around 2019, catalyzed by the 2020 U.S. election cycle. Technology platforms, concerned about the capacity for synthetic media to manipulate users before elections, implemented policies and technical interventions predominantly focused on identifying and labeling synthetic media [7, 44]. Around the same time, technical standards for conveying how content has been generated emerged out of collective attention to synthetic media's risks; in 2020, Adobe, Microsoft, Twitter, and The New York Times spearheaded an initiative for adding signals to visual media via signed metadata, thereby providing audiences with context about visual content, and that has since evolved into an open standard effort called the Coalition for Content Provenance and Authenticity (C2PA) [46].

Most recently, government regulators and standards bodies have begun to act on synthetic media—with clear emphasis on interventions requiring transparency about where and how content was created or edited. In July 2023, the U.S. White House published voluntary commitments signed by many of the largest AI companies, anchored on technical and audience-facing signals for conveying that visual content has been AI-generated [21]. The next year, the U.S. AI Safety Institute launched one of its five working groups on the challenges synthetic content poses to information; it's working to "[identify] the existing standards, tools, methods, and practices...for authenticating content and tracking its provenance, [and] labeling synthetic content" [8]. The EU AI Act also has specific transparency requirements for labeling synthetic media, alongside categorization of types of synthetic media like art, satire, and research that are not considered harmful [34]. There is also rulemaking targeting specific domains touched by synthetic media, ranging from election advertisements, to pornography, to the ways news can be used to train AI models [56, 57]. While there are many societal and ethical implications of synthetic media, here, I focus mainly on one of the most common governance goals: supporting evaluation of media for a variety of aims. Focusing on cross-cutting governance solutions can help illuminate the complexity inherent in the technology's impact—and reveal broader insights into how stakeholders can govern a technology with such multitudinous effects.

### 1.2 Multistakeholder AI Governance

Many have advocated for multistakeholder governance as the optimal way to assess and respond to synthetic media's varied impacts. Here, I borrow the definition of a multistakeholder process from the Internet Corporation for Assigned Names and Numbers (ICANN) as, "one in which there are many stakeholders working together in a learning process and towards a common goal, the work involves different sectors and scale, they deal with structural changes and seek to bring about change, agreements are based on cooperation, stakeholders consciously deal with power and conflict, and there are both bottom-up and top-down strategies integrated into governance and policymaking" [24].

In a landscape review of the AI principles spawned by intergovernmental and multistakeholder organizations since 2016, Fjeld and colleagues [15] suggested that "multistakeholder participation is touted as a mechanism for building normative consensus on the governance of AI technologies and being a policy vehicle for operationalizing such principles." This sentiment is echoed by many institutions in the AI space—ranging from philanthropy to industry to civil society [36, 33]. Evaluating the throughline between multistakeholder collaboration and meaningful AI governance, though difficult, is vital for ensuring such activities are optimally beneficial and that we can draw connections between "governance modalities and governance functions" [42]. More specifically, studying multistakeholder governance efforts focused on synthetic media can shed light on safeguards that help audiences understand content in the AI age.

The International Network of AI Safety Institutes' inaugural meeting, bringing together many countries spearheading work on synthetic content in November 2024, embraced this premise. The synthetic content session highlighted that "multiple stakeholders are often involved in the creation, modification, publication, and dissemination of digital content, and each has a role to play in managing risks" [25]. Indeed, multistakeholder governance has been integral to synthetic media, even before the recent regulatory fervor. Partnership on AI (PAI), a global multistakeholder nonprofit focused on responsible AI development and deployment that has been working on synthetic media governance since 2018 has embraced such a model. While their work has focused on technical and policy guidance that is voluntary in nature, much of their focus is impacting, like AISI, those "creating, distributing, and building tools and technologies" for synthetic media, while also integrating perspectives from other sectors [22, 40].

## 2 MOTIVATION

Despite the growing attention on synthetic media's impacts and uptick in cross-sector governance activities, there has been little empirical focus on what they reveal: about the challenge synthetic media poses, potential solutions, and how those solutions can be enacted in practice. This paper responds to this gap by studying the stakeholders involved in real world examples of multistakeholder synthetic media governance. Focusing on the diverse actors who have been involved in synthetic media governance can not only shed light on safeguards that help people make sense of media today, but it can also contribute to the dearth of first-hand scholarship on the role of multistakeholder work in AI governance that seeks to inform technology practice and product. Specifically, this paper seeks to answer the following research questions:

- RQ1: What do collaborations between civil society, media, industry, and policy stakeholders reveal about synthetic media governance that supports audience understanding of media in the AI age?
- RQ2: How can diverse stakeholders collaborate to implement the noted synthetic media solutions in practice?

## 3 METHODS

My qualitative methodology featured in-depth interviews inspired by real-world examples of synthetic media governance. Anthropological inquiry into AI is a burgeoning field, and it can reveal sociotechnical dynamics of novel AI systems [17, 41]. Further, qualitative research is useful for studying multistakeholder AI governance since it exposes contextual conditions relevant to the domains being studied, simultaneously offering "concreteness and circumstantial specific" and "generalizability" [59, 4, 50, 5]. Three main research phases took place: 1) autoethnographic review of the two case studies in synthetic media governance, 2) interviews with 13 media, industry, and civil society stakeholders who participated in the cases, and 3) interviews with 10 synthetic media government policymakers who did not participate in the cases. In the sections that follow, I motivate the use of PAI cases as the foundation for this research before describing study participants, interview questions, and analysis methods.

### 3.1 Positioning the Partnership on AI (PAI) as a Research Foundation

PAI is a useful venue for investigating multistakeholder work in responsible AI broadly, and for synthetic media specifically. Despite its recent founding in 2016, PAI is one of the most mature and institutionally diverse organizations in the expanding responsible AI space, making it a useful lens for studying stakeholders and practices related to AI governance [15]. PAI has been working on developing cross-sector governance for synthetic media since 2018—providing a useful testbed to interrogate the ways synthetic media can be governed with stakeholders from various sectors. PAI was notably founded by Amazon, Facebook, Google, DeepMind, Microsoft, and IBM, with Apple joining in 2017; however, its board features a multistakeholder balance of public benefit entities and industry representation [35]. PAI has over 120 partner organizations, ranging from the ACLU to OpenAI, and while largely concentrated in North America, has African, European, Asian, and South American partners. From its founding, the organization has focused on bolstering civil society voices. While the organization is based on a model in which for-profit partners contribute charitable donations, public benefit entities do not pay any dues—and the organization also receives a growing share of its funding from philanthropy [39]. Some have expressed frustration with PAI's proximity to industry [1]. However, since its inception most of its partners are civil society organizations, many of which actively *seek* collaboration with industry to advance their work and bolster its technical merit. This tension—the role of industry involvement—is typical of most multistakeholder initiatives that aim to navigate power dynamics. Rather than undermining the organization's value as a foundation for governance research, this dynamic enhances it, offering a rich basis for inquiry and critical analysis [9].

#### *3.1.1Two Synthetic Media Governance Examples: The Deepfake Detection Challenge (Case 1) and Synthetic Media Framework (Case 2)*

Here, I focus on two distinct examples of synthetic media governance from PAI to frame analysis, select participants, and inform interview questions: PAI's governance of Facebook's Deepfake Detection Challenge (Case 1) and PAI's Synthetic

Media Framework (Case 2) [12, 29, 40, 22]. Case 1 focuses on governance of a machine learning competition for algorithms that detect synthetic media while Case 2 is a voluntary synthetic media policy with guidance for those creating, distributing, and building synthetic media (specifically focused on transparency, consent, and responsible and harmful use cases). 18 institutions formally participated in at least one of these examples, and they include actors from media (BBC, CBC, The New York Times), industry (Adobe, Bumble, Microsoft, TikTok, Synthesia, D-ID, Google, Meta, OpenAI, Respeecher), and civil society (First Draft, WITNESS, Meedan, Code for Africa, Thorn, Stanford's HAI).

### 3.2 Interview Participants and Sampling

23 individuals were interviewed in 60-minute in-depth, semi-structured interviews [23]. 13 represented one of the civil society, industry, or media institutions that participated in the PAI cases, and the 10 additional interviewees were government stakeholders who have worked on synthetic media (Table 1). Supplementing the PAI interviews with policymakers ensured there would be participants who may be more critical of PAI and other nongovernmental efforts; inclusion could also help unravel the connections between voluntary guidance like that at PAI and government regulation. While policymakers could be drawn from a global sample, I chose to focus on those in Europe and North America. Those jurisdictions have been most closely coupled with the PAI work and both have profound influence in the global AI governance space [13]. In Section 6, I explore tradeoffs from this decision in more depth.

#### *3.2.1 PAI Stakeholder Participants*

A purposive sampling technique was used to select 13 individuals from the institutions that participated in at least one of the PAI cases. Recruitment priority was given to those who participated in both cases, and several stakeholders declined being interviewed due to employer constraints. This resulted in a total of three civil society, seven industry, and three media participants; they represented a variety of functions within their sectors, ranging across technical and social impact expertise, and roles in AI research, engineering, product, policy, news, and advocacy.

#### *3.2.2 Policymaker Participants*

Ten policymakers were recruited through a snowball sampling technique, ultimately including six from North America and four from Europe. Like the stakeholders in the PAI interviews, policymakers came from a variety of backgrounds ranging from law to politics to research and worked on an array of governance tactics, including legislation and standards.

Table 1: Distribution of Interviewees and Sector

| Sector | Respondents |
|---|---|
| Civil Society | PAI1, PAI3, PAI10 |
| Industry | PAI2, PAI4, PAI6, PAI8, PAI11, PAI12, PAI13 |
| Media | PAI5, PAI7, PAI 9 |
| Government Policy | POL1 - POL10 |

### 3.3 Interview Protocol

All interview questions (see Appendix A) were drawn from themes that emerged during autoethnographic reflection on the PAI cases. For PAI stakeholders, interviews explored what they learned about the field and multistakeholder AI governance through the cases, how sociotechnical collaboration informed their work, challenges in synthetic media, synthetic media's dual use, and interventions for transparency. Policymaker interviews focused on insights on synthetic media and cross-sector governance, challenges in synthetic media, dual use, interventions for transparency and disclosure, and their understanding of governance instruments. The semi-structured nature of the interviews enabled flexibility to dig deeper into interesting strands that arose throughout the interview. Additionally, the researcher is a practitioner in the AI field, which enabled a degree of interviewee openness and transparency about the nuances of their work.

### 3.4 Analysis Methods

I first conducted an informal thematic coding process for autoethnographic reflection on the PAI cases. I then used a form of attribute coding [48] for interviews with four parameters: organization, region, which PAI case they participated in (if not policymakers), and whether they had social impact, technical, or sociotechnical expertise. Preliminary, descriptive codes emerged when going over interview transcripts while listening to interviews, followed by more formal inductive coding of the interview transcripts. I iteratively assigned codes to interview segments into higher level themes for analysis.

## 4 FINDINGS AND ANALYSIS

I first describe high-level, autoethnographic insights from the PAI cases, as those informed the interviews. I then present analysis for three areas affecting synthetic media governance: (1) the challenges synthetic media poses, (2) solutions for addressing those challenges, and (3) how to *implement* those solutions, in practice. Figure 1 illustrates the organizational logic of the findings and the discussion they inform (in Section 5).

### 4.1 High-Level Themes from PAI Cases

I worked firsthand on the PAI cases and began by reviewing them and selecting key themes from existing material. Doing so thematically orients this research and supports testing such assumptions with stakeholders through interviews. Key themes included synthetic media's dual use, the need for sociotechnical interventions on content, the simultaneous value (and difficulty) of supporting transparency and disclosure about content, and the ways in which different actors in the synthetic media pipeline have roles to play in mitigating the harms from synthetic media and optimizing benefits. In addition, it became clear that harmful and responsible impacts of synthetic media do not require AI to manifest; one can deceive, defame, and harass, for instance, *without* AI. Rather, synthetic media is one mechanism affecting existing harms and prosocial dynamics, and lower tech, non-AI methods for manipulating content may be just as harmful as those that rely upon AI. Governance interventions must therefore be responsive to manipulations or edits with non-AI techniques and to shifts in broader societal trust in an era where skepticism about the *prospect* of AI-generated content can make it harder to verify authentic content. PAI cases also highlighted three distinct but related aspects of synthetic media governance that scaffold my analysis: 1) ecosystem understanding of the problem such technology poses, 2) development of norms and solutions, and 3) tactics for implementing those norms and solutions.

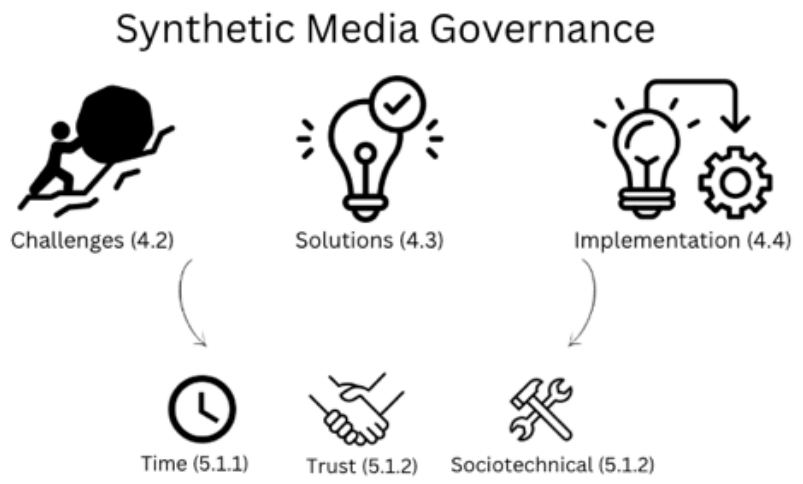

Figure 1: A diagram of the organizational logic of analysis and discussion, made with graphics from Canva.

### 4.2 How Stakeholders Understand the Synthetic Media Challenge

The descriptive and inductive coding processes revealed many shared perspectives on how stakeholders understand synthetic media's challenges. Interviewees emphasized myriad policy questions encapsulated in synthetic media governance and those that unify them, how they grasped dual use to inform policymaking, and the ways in which their sense of past, present, and future informed their conceptualization of synthetic media.

#### *4.2.1 Synthetic Media as a Misrepresentation and Control Challenge*

One policymaker [POL9] summed up the complexity of the synthetic media challenge, suggesting, "Synthetic media is tricky because the policy question is multiple questions." Indeed, throughout the interviews, a wide array of policy challenges emerged pertaining to copyright, privacy, freedom of expression, intellectual property, and transparency, touching domains like child safety, human rights, news, security, and elections. Yet, participants offered ways to coherently think about synthetic media's challenges to support its governance. Several emphasized that they construe the synthetic media challenge related to its **inputs** and **outputs**: the ethics around the data that fuels development of models for generating content with AI and those around how content that emerges from tools and models can be identified and understood, respectively. PAI6 emphasized that they interpret the synthetic media challenge as one related most broadly to questions of **control** of two features: one's likeness and knowledge about where content comes from. Relatedly, stakeholders expressed a shared understanding that the primary challenges posed by synthetic media are centered on **misrepresentation** and **impersonation.** As one media stakeholder described, "There's nothing inherently wrong with synthetic media. What's wrong is misrepresentation. You know, I don't really believe that there's spaceships flying to alternate moons and having starship battles, but they're kind of fun to watch" [PAI9]. Misrepresentation may not be harmful in certain contexts—like movies—and the real concern may be deceptive misrepresentation. PAI2, an industry stakeholder, distinguished between synthetic media and text to highlight synthetic media's implications on misrepresentation:

> *"If you're just reading a text, text is already something that's like…not a direct representation of reality. Text is always conveying information that isn't seen or heard or is like coming from a different place. And then, like, the text is transmitting that information, but it's not information itself. But if you see a photo, that is a representation, we all think that's a representation of reality…The harm of text generation is behavioral, it's how easy it makes it to generate huge amounts of propaganda and to spread it."*

There was consensus among participants that people interpret visuals as records of reality and that AI presents a challenge to this interpretation—distinct from text. Many of those interviewed spoke of broader concerns that synthetic media's impact on misrepresentation and impersonation—even when used for good—could implicate large societal dynamics like epistemic doubt and distrust in information, beyond that of existing manipulation methods like Photoshop. As POL6 stated, as soon as "you can fake everything and you can fake everyone, and you can invent people and do that at scale, the whole setup of the digital space changes." In summary, stakeholders emphasized a focus on misrepresentation as the key challenge synthetic media poses—shifting trust, authenticity, and the very foundation of digital communication.

#### *4.2.2 Synthetic Media as a Dual Use Technology*

Deeper insight into how stakeholders assess synthetic media's dual use can support exploration of governance solutions. When asked about how they evaluate synthetic media's impact, 10 stakeholders (two civil society, one media, two industry, five policy) said it was net negative, four stakeholders (one media, two industry, two policy) said it was net positive, and eight either said it was too complicated to even offer an answer or suggested this was not a productive question to consider. Several spoke of the technology's greater negative impact but tempered their descriptions by stating that it's due to where they're coming from, that they have that opinion "because [they've] had a harm centric mindset in many of the roles [they've] had" [POL3]. Many of those who spoke of positive impacts pointed to lack of data about influential harmful impacts, stating things like, "I feel like synthetic media hasn't been the big threat that we thought" [POL1] and how their pragmatism about the technology's inevitability requires them to channel it for good. Of those who were reluctant to answer, most expressed their uncertainty from a place of humility, emphasizing that it is impossible to measure and evaluate all the varied impacts of a single technology and opine on its relative harm. One individual seemed upset that the question was being asked, expressing conviction that even assessing dual use has no bearing on how the technology is governed, which should be oriented around specific impacts rather than the technology broadly.

A unique feature about synthetic media's dual use emerged through data analysis. Participants discussed how harmful use cases can be seen well before most positive use cases, largely since harmful use cases do not need technology to be perfectly robust or accurate to have their intended negative impacts. As PAI1 elaborated, "For malicious uses...technical improvements make them easier, and the technical failures don't really make it any less useful." They went on to describe how areas that require reliable information are "reliant much more on concepts we haven't really landed." POL6 suggested that "even GPT2 was good enough" to cause harm. Contrastingly, "for the beneficial uses, if you say we can finally get, I don't know, a chatbot that gives people a port of call for medical questions or we can revolutionize so many different fields, the quality isn't there" [POL6]. This premise, that negative use cases materialize more quickly and require less sophisticated versions of the technology, affects many of the stakeholders' present-day assessment of synthetic media's impacts, which form the foundation for how they develop synthetic media governance.

#### *4.2.3 Temporal Perspectives Inform Synthetic Media Challenges*

Like the observation that positive use cases take longer to develop, many other stakeholder insights into synthetic media's challenges were tied to their temporal perspectives on the past, present, and future. More than half of participants explicitly referenced synthetic media's challenges as part of a continuum, rather than completely novel challenges. Participants described the fact that "people were manipulating photos decades ago, probably for as long as photos exist" [POL6] and how they "always talk about this as an evolution of the technology" [PAI10], how "in many ways this is not novel, it's like an accelerated problem" and that the "conundrum of trustworthiness [of information]" has "existed since the beginning of the Internet" [POL3]. Many participants went on to describe how what makes the current synthetic media challenge different from previous challenges around manipulation of visuals is related to the time it takes to create images today—another temporal detail. Others noted how the time that it takes to create synthetic media, and share it, can affect longstanding processes like image verification and child safety triage, making it vital to respond to this change at a faster pace compared to the past. As POL3 stated, "Speed is often of the essence in these kinds of situations. If it takes you four days to verify something as a deepfake…by then the news cycles moved on…the goal has already been achieved and the pace you're able to respond to these things is harmed." The time required to verify content, the perception of synthetic media as part of an evolution from a long history of visual manipulation, and the recognition that positive use cases will take longer to develop than negative ones shape how stakeholders understand governance challenges and their proposed solutions.

### 4.3 How Stakeholders Understand Synthetic Media Solutions

In this section, I analyze the solutions participants proposed for challenges outlined in Section 4.2. They highlighted both the limitations and potential of technical approaches for identifying and disclosing synthetic media, as well as the pivotal role of trust for ensuring effective media transparency solutions.

#### *4.3.1 Technical solutions will not solve synthetic media challenges, but they are still worthwhile.*

Several technical governance proposals for identifying and describing synthetic content emerged throughout interviews, specifically deepfake detection and content provenance methods. Unlike deepfake detection, content provenance methods like watermarks, fingerprints, and cryptographically signed metadata are purposefully applied to media and can be identified by third parties who are able to interpret the signal [30]. This is a common solution, one that has been explored by C2PA, suggested in PAI's Synthetic Media Framework, pursued by many builders, creators, and distributors of synthetic media, and even proposed in legislation like California's AB-3211 on Digital Content Provenance Standards.

Stakeholders were committed to these technical transparency solutions despite highlighting their limitations. An industry stakeholder [PAI13] said they were surprised to see how limited watermarking was for supporting identification of synthetic media. A civil society stakeholder [PAI10] expressed surprise at how imperfect detection is and how little progress has been made in improving it since Case 1. A media stakeholder talked about how their takeaway from Case 1 was that it would be "unethical to deploy deepfake detectors," alluding to the adversarial dynamics that make detectors susceptible to evasion [32]. And yet, despite these limitations, the stakeholders underscored their hope that such technical interventions can be improved through sociotechnical collaboration and their conviction that detection and provenance are the best options we have to date and must therefore be pursued. However, even if such methods could consistently and accurately identify AI-edits or generation, merely identifying if content has been edited or generated with AI says nothing of whether media features harmful misrepresentation, which is the most vital challenge to respond to and described in Section 4.2.

#### *4.3.2 Transparency must move beyond the "AI or not" binary, to provide meaningful, broad context.*

Stakeholders supported solutions that shed light on the type, degree, and impact of manipulation or synthesis more than those that solely emphasized AI's presence. As POL3 said, "synthetic or not is not the right question because it forces people to think in a false binary, instead of helping people think if I should trust this content, which the 'synthetic or not' piece is one teeny subset." Others echoed this point, invoking dual use, stating "AI generated content is not inherently bad" [POL2] and "we really need to be careful about how we think about AI as inherently sinister" [PAI1]. Others provided examples of AI-manipulations which did not automatically contribute to misrepresentation, like color correction. While many acknowledged the limits of providing cues that *solely* indicate whether AI has touched a piece of media, they generally supported details about how AI was used to edit or generate imagery as *one of several* signals that could support a viewer's media understanding.

Two specific solutions emerged on how to move beyond the AI or not binary. One focused on differentiating mechanical and human-made AI changes, and another suggested relying on the news industry's editorial standards. They suggested that this would better differentiate AI tweaks that are immaterial to artifacts serving as records of reality, by focusing on material changes (in the case of newsrooms standards), and using "mechanical" as a proxy for immaterial AI changes that happen due to the natural flow of media through the information environment (in the mechanical vs. human example). Indeed, many stakeholders recognized a pressing need to move beyond the "AI or not" binary, since they suggested all content will soon feature some AI elements.

After emphasizing the limitations of using AI's presence as a proxy for misrepresentation, many participants stressed the importance of advancing current approaches to audience-facing labels for synthetic media. PAI3 noted that their key takeaway from Case 1 was the limit of labels: the challenge of providing the right context to support informed decision making and encouraging broader public engagement with media transparency. An industry stakeholder described how they are "coming to terms with how limited labeling and disclosure is" and how with so much "hype around AI people just want to see that you've labeled it." Nonetheless, most felt committed to reform, rather than abandon, labels. As POL1 highlighted, "What we consume with our mind should be labeled the way that what we consume with our body is. If it's digested mentally, it's like we digest it physically."

Interviewees were asked how to "meaningfully convey media manipulations to the public," a need PAI emphasized in 2020 [29]. Participants responded by highlighting a need to add context to all media, not just media augmented or generated with AI, and they frequently emphasized media transparency efforts from before the AI era. For instance, some stakeholders emphasized the importance of knowing media's source, with one policymaker describing the "identity of the creators" as "woefully underemphasized" even though it "matters more than anything" [POL5]. Others emphasized a need for greater clarity about edits made to specific regions of interest on visuals, while also recognizing that audience research has shown that people don't care about granular details about content edits and want to understand if content is true or false, real or fake. Of course, this tension—between users benefitting from simple, normative content labels about the *meaning* of content and the desire by those governing the content to avoid normative judgments about content meaning—presents a challenge for synthetic media solutions.

When asked for analogies from other fields to illustrate media transparency best practices, one stakeholder invoked a "recipe" that reveals the complexity and layers involved—some influenced by AI and others not. This approach, they suggested, would help audiences make informed decisions by considering all the elements that shape media. They explained that optimal explanations would describe a "process rather than a product" [PAI1] and defy binary descriptions about the method of creation or editing. Additionally, key features of successful disclosures were referenced, including standardization of visual signals, ensuring audiences trust the authority providing the disclosure, moving beyond a label that relies upon binaries to show nuance and the process of how content was made and/or edited, and simplicity and ease of understanding for audiences who may not be inclined to care about media transparency signals. Ultimately, participants emphasized the limits, but importance, of technical solutions for evaluating and contextualizing media artifacts, with specific focus on the need to identify and disclose more than just AI's presence when describing media.

### 4.4 How Stakeholders Implement and Enact Solutions

PAI1 highlighted three stages of multistakeholder synthetic media governance. The first two have been explored in the earlier sections: 1) developing field-wide understanding of the issue and challenges and, 2) creating norms and guidelines for how to solve to those challenges. Here, I explore the third stage: how stakeholders *implement* those norms and solutions. Pursuing different governance instruments can affect how solutions are enacted, and interviews about implementation shed light on synthetic media governance overall.

*4.4.1 Nongovernmental, voluntary guidelines serve an important, albeit limited, role in synthetic media governance.*

Plausibly, those who participated in the PAI cases had enough faith in the impact of nongovernmental, voluntary standards to work on such efforts; however, several offered more nuance about the role of these instruments when enacting synthetic media solutions. It's important to note that PAI's work is not just distinct from that of government since it is voluntary; governments pursue voluntary standards (e.g., the U.S. National Institute of Standards and Technology or NIST).

PAI's independence was noted as a key feature motivating participation in its work, suggesting that even with financial interests from corporate entities, its degree of independence ensures solutions have impact. As PAI3 stated:

> *"Having an organization that is not in the reporting structure of one of the organizations, one of the stakeholders, one of the interested organizations is very helpful. PAI is ultimately financially interested in Facebook being happy. But there's just enough distance there. And it does understand that if it carries too much water for industry, it doesn't have a place in its identity."*

Also, governance produced at organizations like PAI can move more swiftly to assess governance needs, articulate norms, and test those norms in practice—eventually reaching policymakers when they embark on their own governance. As PAI11 said about voluntary guidelines, they are "an important first step to get convergence…amongst ecosystem actors" and to identify insights that should be "incorporated into legislation." Even policymakers who had not participated directly in voluntary, nongovernmental efforts like the PAI cases emphasized the ways in which well-timed voluntary governance, informed by a cross-sector cohort, informs their work. POL1 stated that having "civil society *and* government push the boundaries on, what are the rules of the road that we need to abide by…is very important." While multistakeholder voluntary guidance is valuable for setting norms and informing regulation, many emphasized that it cannot *replace* actual regulation. Instead, it can only serve as an evidence base to support future regulatory efforts. Describing the limits of NIST's voluntary nature, and their statement that "voluntary commitments are all bogus" POL9 exclaimed, "Companies love NIST…they send them Valentine's Day cards because NIST is voluntary. And if you say I'm complying with NIST guidelines, everyone loves you. Regardless of what you are doing, because there's no enforcement." An industry stakeholder who has worked on voluntary commitments elaborated, "They are not effective at motivating action of any kind, let alone collective action. They are very helpful for understanding the motivations of people who sign up for voluntary actions... but without any teeth, it's very difficult" [PAI13]. For voluntary, nongovernmental instruments to affect synthetic media solutions, it must precede and inform regulation, be independent, center civil society alongside industry, and as some suggested, embolden participants to advocate for governance practices at their own institutions.

*4.4.2 Trust between diverse stakeholders is integral to synthetic media governance, and radical honesty about corporate interests is vital to progress.*

All stakeholders highlighted that synthetic media governance can progress more effectively when there is clear understanding of the incentives driving different institutions and trust between actors. Participants from industry, civil society, media, and policy stressed that corporate incentives are key factors influencing implementation of synthetic media governance; recognizing these incentives and finding ways to address them fosters trust among the diverse groups collaborating to govern the technology effectively. POL2 summed up their understanding bluntly, "My position is whenever we're dealing with regulation or legislation, we're dealing with companies that have two motivations in this world: making money and not breaking the law." Others elaborated on these financial and legal interests, extending them to media companies, not just technology companies, interested in making money and staying relevant in the AI age. A media participant said, "People that control the screens, whether that's the platforms or the broadcasters or everything else, are very, very jealous, of every square inch of real estate on the screen." PAI9 went on to emphasize how this limits any standardized media transparency solutions across entities where audiences encounter media. Others mentioned corporations' "self-interest" [PAI2], "a situation where our shareholders, systems, and incentives are so strong", and their realization of "how loud money is in [synthetic media governance] and how it'll change priorities" [PAI13].

Despite the seemingly straightforward corporate motivations focused on the bottom line, one industry stakeholder expressed surprise at how challenging it is to achieve even minimal industry consensus. They noted that stakeholders tend to be "defensive" about their perspectives, working to "preserve [their] capitalistic interests" [PAI13]. And yet, most stakeholders—including many policymakers—still saw value working with corporations, alongside uplifted civil society voices, to drive implementation of synthetic media governance. They explained that addressing corporate incentives involves channeling voluntary guidelines into regulation. Additionally, fostering interpersonal trust that goes beyond corporate interests, through open and honest collaborations on a topic from its early stages, can help operationalize guidance and support implementation. As the same media executive who highlighted corporate self-interest stated, "trust happens over time" and when it builds to "sufficient capacity, then you can focus on solving problems, not taking corporate positions." As a civil society advocate suggested, such trust is built through conversations that are "early and ongoing" and that enable "a nuanced conversation that allows people to feel they can stretch their institutional perspective" [PAI1]. Clarifying institutional incentives, while building trust between individuals from those institutions and beyond, ensures that multistakeholder governance can be developed honestly and consider sociotechnical dynamics. Only then can it shape government regulation to respond to corporate incentives.

*4.4.3 Temporal perspectives inform how synthetic media governance is implemented*

Just as temporal perspectives provided a throughline for how stakeholders conceptualized synthetic media challenges, their understanding of time—past, present, and future—also influenced their approach to implementing synthetic media solutions. The rapid advancement of synthetic media specifically, and AI innovations more broadly, led many stakeholders to stress the need for AI governance to be iterative and adaptable. Several pointed out that the technology would look completely different in the future, making it very challenging to create policies that would remain effective over time. One industry stakeholder stated that "locking anybody into a particular approach, a particular technology doesn't make sense. Because no matter what the technology and people's understanding of the technology is now, it will change, and so policies become irrelevant very quickly" [PAI8]. Another policymaker emphasized that "any framing we establish today doesn't necessarily have to be the framing forever; whatever regime was put into place can be amended" [POL2]. The uncertainty about the future justified the need for adaptable, iterative governance.

Several stakeholders discussed how shifts in public consciousness of AI tool use — largely catalyzed by the introduction of ChatGPT by OpenAI in 2022—also changed their sense of how quickly multistakeholder bodies must respond to synthetic media, prompting greater pressure from the press and government, too. This historical moment prompted a sense of urgency around rulemaking for synthetic media, with stakeholders feeling the need to act without complete information or evidence about the technology's impact. In contrast, from 2017 to 2022, stakeholders felt they had more time to "prepare and not panic," as described by WITNESS' program focused on synthetic media [18]. Additionally, this shift in the timescale for creating synthetic media governance affected key features of governance development in the PAI cases and beyond. A civil society stakeholder described a shifting willingness of technology companies to engage in multistakeholder rulemaking on synthetic media compared to the past, partially attributed to the threats of regulation and the connections of much work in synthetic media governance to speech and content moderation challenges made more complicated due to political headwinds. Tactically, as the pace of change in synthetic media policymaking has increased, it has become harder to pursue the "early and ongoing efforts" [PAI1] that are required to blend social impact and technical expertise.

## 5 DISCUSSION

While synthetic media is an emergent visual technology, the pace of development and wide-reaching implications warrant increased attention from AI practitioners and scholars. As cross-sector actors collaborate to inform synthetic media governance, we must study the *people*—specifically decision-makers—shaping such technology development and governance. Policymakers, industry, civil society, and media stakeholders all contribute to the design, development, and deployment of this new visual technology, shaping how the public encounters and makes sense of media in the AI age. This research highlights how these stakeholders govern synthetic media—identifying challenges, solutions, and implementation strategies. Several key themes emerged across the interviews that inform both understanding and enactment of synthetic media governance: 1) the role of stakeholders' sense of past, present, and future, 2) the need for governance efforts to reckon with the limits of technology, and 3) the importance of trust—both between audiences and solutions, and among diverse stakeholders—in supporting governance effectiveness.

### *5.1.1Temporal Perspectives Affect Synthetic Media Governance*

The role of temporal perspectives in stakeholders' understanding of synthetic media and their ability to influence its governance emerged throughout the interviews. On the one hand, most stakeholders emphasized that synthetic media is merely an evolution of an existing problem from the past. Yet, they all expressed a simultaneous sense that the "existing problem"—mainly misrepresentation—is evolving so rapidly, that stakeholders feel they're running out of time to intervene and address it with the urgency it warrants. This is seemingly contradictory: if synthetic media is merely a continuation of existing problems of manipulation, perhaps changing the scale of impact but not type of impact, the need to attend to its governance should be rendered less pressing. Yet this is not what stakeholders expressed, perhaps implying the synthetic media moment *is* a paradigmatic shift for policymaking on misrepresentation.

Indeed, temporality is key to understanding stakeholder governance of synthetic media, as those interviewed demonstrated both reflection on past currents and a sense of urgency about the future of media, information, and interaction. Whether it was a perception that positive use cases will take longer to emerge than negative uses, an expressed need for governance that could adapt over time, or even a recognition that the pace of synthetic media creation makes its impact more profound, stakeholders understood synthetic media governance in time. And solutions, too, were thought about in their temporal context. The nongovernmental, multistakeholder efforts like the PAI cases had impact since they built trust and mapped the field years before the ChatGPT-inspired regulatory momentum of recent years. Timing is vital for getting ahead of regulation, building trust to inform sociotechnical solutions, and responding to future technical innovations and cultural, political, and economic events. While many reduce AI governance through the lens of long-term and near-term risks, thinking more broadly about how temporality affects decision-making amongst those governing AI would support more granular understanding of AI's challenges, solutions, and governance effects [11, 47].

### *5.1.2 Sociotechnical Solutions and Trust*

Ensuring technical solutions are embraced as only one important, albeit limited, part of synthetic media governance complicates many proposed solutions for synthetic media. And multistakeholder collaborations—like those in Case 1 and Case 2—can enrich corporate and policymaker understanding of technical opportunities and limitations. Whether describing the imprecision of deepfake detection, or the robustness of content authenticity signals like watermarks, stakeholders drew attention to how multistakeholder collaboration ensures rulemaking is attentive to where such technical remedies serve a purpose, where they do not, and how they can be improved and supplemented with other non-technical solutions. For instance, several emphasized the need to improve media literacy whenever exploring technical transparency solutions. POL3 explained how, "The technical folks [they've] been seeing have the same qualms, how are we gonna get media literacy and user understanding of this right, which is a fundamental social problem. Like, you could use all the watermarking techniques you want. But if you don't translate that to increase trustworthiness from audiences, then what's the point?" While technology companies may boast about technical solutions for detecting synthetic media or labels that are placed on artifacts, these do not necessarily address the broader, social challenges of digital trust in information. The interviews not only exemplify the limits of technical solutions for synthetic media rulemaking, but also the ways collaborations between diverse stakeholders can best support this sociotechnical empiricism and media literacy work.

When thinking through solutions for synthetic media's challenges, trust emerged as vital prerequisite; trust was not only important *between* the diverse stakeholders working to govern the technology, but also important between audiences and the solutions. Several stakeholders described how to ensure transparency signals are trusted, thereby supporting audience understanding of how content was made and edited, and aspirationally, media literacy. First, they must move beyond just describing "AI or not" to feature broader context, capturing the "recipe" of content creation and alteration. Most agreed that standardizing the visual signals accompanying content would support trust between audiences and signals across surfaces. Further, stakeholders underscored a need for disclosures to be ubiquitous, but not so pervasive that they lead audiences to tune out the content disclosed [49, 58]. While there is no formal consensus on what leads synthetic media solutions to be trusted by audiences, this research offers some best practices and emphasizes the importance of studying this question in the future. Beyond the ideas for synthetic media governance, like labels and disclosures that must be trusted, stakeholders emphasized trust as a vital prerequisite for effective development and implementation of such ideas themselves. In other words—trust *between* the stakeholders governing technology is what enables them to actually collaborate, to move beyond the corporate talking points that can thwart collective rulemaking, and to enrich the specificity of solutions by better accommodating collaboration, consensus, and exploration of how to operationalize themes in practice.

Ultimately, this study untangles how stakeholders governing synthetic media conceptualize, develop, and implement safeguards to help audiences make sense of content in the AI age. Doing so highlights synthetic media as a problem of misrepresentation and impersonation, not simply technology, that warrants a simultaneous understanding of the past, present, and future to govern effectively.

## 6 LIMITATIONS AND FUTURE WORK

By embarking on a broadly ambitious overview of synthetic media governance—its challenges, solutions, and tactics—this research reflects the complexity of the synthetic media landscape. However, this approach sacrifices some depth on each topic. Issues such as how multistakeholder efforts navigate power dynamics, the various implications of synthetic media, and different policy interventions each warrant their own focused research in the future. By analyzing these topics together, throughlines between the real-world processes for governing an emergent technology can emerge, as no single synthetic media stakeholder navigates just one aspect of these themes when pursuing governance.

In addition, this research primarily focused on stakeholders involved in two examples of synthetic media governance led by an independent nonprofit developing voluntary guidance. The stakeholders' familiarity with the PAI cases allowed them to offer detailed critiques of these efforts, but at the same time, their participation in these initiatives suggests an inclination toward the value of multistakeholder approaches to governance. Supplementing with policymakers addressed this limitation, but interviewing a broader range of individuals could provide a more comprehensive perspective. The North American and European focus of this work is also a limitation for making broader claims about synthetic media governance. While North America and Europe have been some of the earliest jurisdictions to work on AI governance, with global dominance in AI policymaking overall and synthetic media policymaking specifically, they present a distinctly Western viewpoint on how AI is governed and do not represent all global governance [13]. Future research should draw from a population of policymakers beyond those two continents and help elucidate the ways in which Western values are dictating the trajectory of AI development and synthetic media solutions.

## ACKNOWLEDGMENTS

Thank you to Dr. Kathryn Eccles for her guidance and kindness as I developed this research and for feedback on late-stage drafts. I am grateful for the support of the Oxford Internet Institute Shirley Scholarship and Dieter Schwarz Fellowship for enabling me to pursue my research, and for time and space at the Rockefeller Foundation's Bellagio Residency to develop the idea for this project. Additionally, while this work is distinct from my role at Partnership on AI, I could not pursue such independent scholarship without the steadfast support of my colleagues, Dr. Stephanie Bell and Rebecca Finlay, who are themselves committed to evidence-based policymaking.

## COMPETING INTERESTS

I am employed full time at Partnership on AI while pursuing my doctorate at Oxford. Partnership on AI is funded by a combination of philanthropic institutions and corporate charitable contributions. Primary corporate funding is always considered general operating support and legally classified as non-earmarked charitable contributions (not donations in exchange for goods or services, or quid pro quo contributions) to avoid the possibility of conflict in corporate funders having undue influence on Partnership on AI's agenda or on particular programs. More detail on Partnership on AI's funding and governance is available online. While this document reflects the input of individuals representing many PAI Partner organizations, it should not be read as representing the views of any organization or individual or any specific PAI Partner.

## POSITIONALITY STATEMENT

It is important to position myself as the researcher in this study, with direct, first-hand experience in the case studies described and professional relationships with several of the stakeholders interviewed. I have worked at PAI since the organization's inception and led both efforts described in this paper. Notably, this project is distinct from my work responsibilities at PAI, involved no editorial oversight from the organization, and was conducted in my own time. Borrowing from Crawley's (2014) approach, I take inspiration from autoethnography, affording me a process of vigorous self-reflection—one that enabled me to engage with my positionality critically and candidly in this work.

To some degree, my proximity to the activities of the cases posed a challenge; clearly, as is made apparent from my career choices, I support the premise and promise of multistakeholder work. Further, I have professional relationships with many of those I interviewed, and some may worry that made me less likely to ask difficult questions. Additionally, some may think that I would fail to be meaningfully critical of an AI governance effort at my own institution, to market the organization as doing good work; however, this was not the case, and I argue that proximity offered a greater volume of detail available for analysis, probing, and critique, and that proximity does not preclude meaningful introspection and criticism. A failure mode of this research was that this piece merely reads like an impact story to market an organization like PAI, rather than a critical reflection on how PAI's work can provide a glimpse into synthetic media governance. Throughout, I engaged critically with participants, expanded the sample of interviewees to include those outside my professional milieu, and openly described critiques of multistakeholder governance like PAI's.

Further, this work is distinct from my day job in the eyes of PAI and the OII, and participants I interfaced with seemed more likely to be forthright with me since I am myself a practitioner in the AI and media integrity community—working at a neutral third-party nonprofit organization, where I have a great deal of autonomy and freedom to write critically. Additionally, the Oxford Central University Research Ethics Committee approved my positionality statement alongside my research ethics review before the project was conducted

## ETHICAL CONSIDERATIONS STATEMENT

This research was approved by the University of Oxford's Central University Research Ethics Committee with reference number OII_C1A_24_004.

# A APPENDICES

## A.1 PARTNERSHIP ON AI AND POLICYMAKER: INTERVIEW QUESTIONS

**For those in civil society, industry, media:**

**Introduction**

Introduce self. Reiterate details from information sheet and receive informed consent orally.

**Background**

1 — Please share a high-level description of your disciplinary background as it relates to AI governance.

2 — Broadly, what is your role in synthetic media governance?

Case Studies

3 — What key insights on synthetic media emerged for you during your involvement in Case 1 and/or Case 2?

4 — What key insights on multistakeholder AI governance emerged for you during your involvement in Case 1 and/or Case 2?

5 — How must social and technical expertise be brought to bear and balanced when thinking about synthetic media governance?

**Key Challenges in Synthetic Media**

6 - What do you describe as an ethical or responsible way to develop, create, and share synthetic media?

7 — Dual Use: How do you understand the balance of synthetic media's positive and harmful impacts? How do we govern the technology to balance these as such?

8 — Transparency and Disclosure: What do the example cases of PAI governance reveal about open questions and opportunities related to disclosing and relaying that content has been synthesized for audiences? Possible follow up: How exactly should we label AI-generated media for audiences?

9 — What barriers must be overcome for diverse stakeholders to support audience encounters with accurate, and not harmful, synthetic media?

**Closing**

10 — Is there anything else I did not ask that you think is related to the topic of multistakeholder synthetic media governance?

---

**For those in policy:**

**Introduction**

Introduce self. Reiterate details from information sheet and receive informed consent orally.

**Background**

1 — Please share a high-level description of your disciplinary background and expertise.

2 — Broadly, what is your role in synthetic media governance?

**AI Governance Methods**

3 — What key insights on synthetic media have emerged for you during your work in AI policy?

4 — How have cross-sector perspectives from civil society, industry, and media (and potentially others, like the public) informed your policymaking?

5 — How must social and technical expertise be brought to bear and balanced when thinking about synthetic media governance?

**Key Challenges in Synthetic Media**

6 - What do you describe as an ethical or responsible way to develop, create, and share synthetic media?

7 — **Dual Use**: How do you understand the balance of synthetic media's positive and harmful impacts? How do we govern the technology to balance these as such?

8 — **Transparency and Disclosure:** How do you understand policy's role in supporting the disclosure and communication that content has been synthesized using AI? Possible follow up: How exactly should AI-generated audio-visual content be labeled for audiences?

9 — What barriers must be overcome for diverse stakeholders to support audience encounters with accurate, and not harmful, synthetic media?

10 — What key tensions must the policy community resolve to effectively govern synthetic media? For example, expression and safety.

**Closing**

10 — Is there anything else I did not ask that you think is related to the topic of government regulation of synthetic media?